\documentclass[authoryear]{elsarticle}
\usepackage{setspace}   %Allows double spacing with the \doublespacing command
\usepackage{rotating}
\usepackage{float}
\usepackage{hyperref}
\usepackage{amsmath}
\usepackage{amssymb}
\usepackage{color}
\newcommand{\beginsupplement}{%
        \setcounter{figure}{0}
        \renewcommand{\thefigure}{S\arabic{figure}}%
     }

\begin{document}
\begin{frontmatter}
\begin{abstract}
\noindent We investigate the relationship of resting-state fMRI functional connectivity estimated over long periods of time with time-varying functional connectivity estimated over shorter time intervals. We show that using Pearson's correlation to estimate functional connectivity implies that the range of fluctuations of functional connections over short time scales is subject to statistical constraints imposed by their connectivity strength over longer scales. We present a method for estimating time-varying functional connectivity that is designed to mitigate this issue and allows us to identify episodes where functional connections are unexpectedly strong or weak. We apply this method to data recorded from $N=80$ participants, and show that the number of unexpectedly strong/weak connections fluctuates over time, and that these variations coincide with intermittent periods of high and low modularity in time-varying functional connectivity. We also find that during periods of relative quiescence regions associated with default mode network tend to join communities with attentional, control, and primary sensory systems. In contrast, during periods where many connections are unexpectedly strong/weak, default mode regions dissociate and form distinct modules. Finally, we go on to show that, while all functional connections can at times manifest stronger (more positively correlated) or weaker (more negatively correlated) than expected, a small number of connections, mostly within the visual and somatomotor networks, do so a disproportional number of times. Our statistical approach allows the detection of functional connections that fluctuate more or less than expected based on their long-time averages and may be of use in future studies characterizing the spatio-temporal patterns of time-varying functional connectivity.
\end{abstract}

\title{Dynamic fluctuations coincide with periods of high and low modularity in resting-state functional brain networks}
\author{
Richard F. Betzel$^{1,2,*}\corref{corr}$,
Makoto Fukushima$^2$,
Ye He$^4$,
Xi-Nian Zuo$^4$,
Olaf Sporns$^{2,3}$
}
\address{$^1$ University of Pennsylvania, School of Engineering and Applied Science, Department of Bioengineering, Philadelphia PA, 19104, USA}
\address{$^2$ Indiana University, Psychological and Brain Sciences, Bloomington IN, 47405, USA}
\address{$^3$ Indiana University, Network Science Institute, Bloomington IN, 47405, USA}
\address{$^4$ Key Laboratory of Behavioral Science and Magnetic Resonance Imaging Research Center, Institute of Psychology, Chinese Academy of Sciences, Beijing, China}

\cortext[corr]{corresponding author: \texttt{rbetzel@seas.upenn.edu}}

\begin{keyword}
Networks, Dynamic functional connectivity, Modularity
\end{keyword}

\end{frontmatter}

%\linenumbers
\section*{Introduction}
\doublespacing
The human brain can be described in terms of a network of neural elements and their connections \citep{bullmore2009complex}. In cases where the network refers to structural connectivity, the connections indicate the presence of an anatomical link, such as a synapse between two neurons or a white matter fascicle between brain regions, the complete set of which defines the \textit{human connectome} \citep{sporns2005human, sporns2011networks}. In cases where the network refers to functional connectivity, a connection indicates that a pair of neural elements is functionally related to one another, most commonly expressed through estimating statistical dependencies among neuronal time courses  \citep{friston2011functional}. The complete set of functional connections between all pairs of neural elements specifies a \textit{functional brain network}. In most current applications, functional brain networks are constructed by calculating Pearson's correlation coefficient of blood-oxygen-level-dependent (BOLD) time series for all pairs of ROIs, recorded over an extended period of time during rest or task conditions. The result is a full matrix, whose coefficients represent estimates of the magnitude of the functional connection between each pair of ROIs.

In calculating the magnitude of a functional connection, one typically takes into account BOLD fluctuations over an extended  scan session lasting on the order of 5-10 minutes or longer. The matrix of correlations generated for the entire scan session is here referred to as the \textit{static functional connectivity} matrix. Recently, it has been proposed that such an approach overlooks potentially meaningful fluctuations in the magnitude of functional connections that take place on shorter time scales \citep{chang2010time,hutchison2013dynamic, calhoun2014chronnectome, kopell2014beyond}. A common approach to obtain an estimate of this so-called \textit{dynamic functional connectivity} is to divide the scan session into (potentially overlapping) sub-intervals or windows, and calculate a full correlation matrix for each sub-interval. The correlation matrix for each sub-interval is taken as an estimate of the instantaneous functional connectivity. This \textit{sliding window} approach has been applied widely to study dynamic connectivity of functional brain networks estimated from both fMRI BOLD \citep{allen2012tracking, bassett2013task, hutchison2013resting, gonzalez2014spatial, zalesky2014time, shen2015stable, shen2015network, karahanouglu2015transient} and electrophysiological data \citep{kramer2011emergence, chu2012emergence, betzel2012synchronization, doron2012dynamic, tagliazucchi2012dynamic, de2012cortical, chang2013eeg}.

Despite its rapid adoption, the methodological pitfalls of using such an approach to obtain estimates of dynamic functional connectivity are not fully understood \citep{hlinka2015danger, zalesky2015towards, leonardi2015spurious}. One of the most important and technically challenging aspects of dynamic functional connectivity analysis is determining whether observed fluctuations in a functional connection over time are meaningful, both in terms of statistical significance and neurobiological function \citep{liao2014dynamicbc, lindquist2014evaluating, thompson2015mean}. 

Another important topic in network neuroscience is the concept of modular brain networks \citep{sporns2015modular}. Modular networks can be sub-divided into nearly autonomous sub-systems, also known as modules or communities. In the context of functional brain networks, communities have spatial distributions similar to the brain's functional systems \citep{power2011functional, yeo2011organization} as well as distinct behavioral and cognitive fingerprints \citep{crossley2013cognitive}. An important question concerns the timescales over which these functional systems emerge \citep{laumann2015functional} and the extent to which they fluctuate over shorter timescales \citep{bassett2011dynamic, bassett2013task}.

In this report, we show that the strength of static functional connectivity imposes constraints and implies statistical expectations on the range of fluctuations expressed in dynamic functional connectivity. We propose a non-parametric, model-free statistical test for identifying unexpected fluctuations in functional connectivity over time. We then apply this technique to fMRI data acquired from $N=80$ individuals and identify intermittent episodes in resting brain activity when functional brain network exhibit varying levels of fluctuating dynamics. Episodes of high modularity characterized by an over-expression of stronger/weaker than expected functional connections, are interspersed among periods of low modularity during which connection weights are within the statistical limits predicted by their static connectivity levels. Finally, we find that the stronger/weaker-than-expected connections show specific spatial distributions, and include disproportionately many connections within and between visual and somotomotor systems.

\section*{Methods}

%% this needs to be changed -- it's taken directly from Betzel et al 2015
\subsection*{Image acquisition and processing}
All data analyzed in this report come from the NKI-Rockland Lifespan Sample \citep{nooner2012nki}. This study was approved by the NKI review board and all participants provided informed consent prior to data collection. As part of the data collection process, each participant completed one anatomical scan, one diffusion structural scan and three resting-state functional MRI (rfMRI) scans that varied in terms of TR time, voxel size, and scan duration: 1) TR = 2,500 ms, voxel size = 3 mm, scan duration = 5 min; 2) TR = 1,400 ms, voxel size = 2 mm, scan duration = 10 min; and 3) TR = 645 ms, voxel size = 3 mm and duration = 10 min. We analyzed the fastest multiband imaging data, which appeared superior to the other acquisitions in terms of both accuracy and reliability of rfMRI \citep{zuo2014test}. More details on these data are publicly accessible via the FCP/INDI website
\url{(http://fcon_1000.projects.nitrc.org/indi/enhanced/index.html)}. All image data were preprocessed using the Connectome Computation System (CCS) pipeline. The preprocessing strategy included discarding the first several volumes (10 seconds), removing and interpolating spikes that arise from either hardware instability or head motion, slice-time correction, image intensity normalization, and removing the effect of physiological noise by regressing out twenty-four parameters from a motion model \citep{yan2013comprehensive, satterthwaite2013improved} as well as nuisance variables such as white matter and cerebrospinal fluid signals, along with both linear and quadratic trends. Details of the image preprocessing steps are described in  \cite{xu2015connectome}. In total, we processed data from 418 individual participants. The quality control procedure in the CCS excluded 19 participants due to their low-quality multimodal imaging datasets, which met at least one of the following criteria: (1) failed visual inspection of anatomical images and surfaces; (2) mean frame-wise displacement $>$ 0.2 mm; (3) maximum translation $>$ 3 mm; (4) maximum rotation $>$ 3$^{\circ}$; or (5) minimum cost of boundary-based registration (a measure of image registration quality) $>$ 0.6. Additionally, 51 participants were excluded from subsequent analyses because of clinical diagnoses as defined by DSM-IV or ICD10 and 32 due to incompleteness of the multimodal imaging datasets. This leads to a lifespan sample of 316 healthy participants. Within this sample we focus on a sub-sample of healthy adults aged 18-30 years, comprising 80 participants.

\subsection*{Static functional connectivity}
We term the functional connectivity network estimated over the duration of the entire scan session as the \textit{static functional connectivity} (sFC). We constructed for each participant a sFC matrix, $\mathbf{W}$, whose elements were equal to $W_{ij}=\frac{1}{T-1}\sum_{t=1}^T z_i(t) \cdot z_j(t)$, where $\mathbf{z}_i = \{z_i(1),\ldots,z_i(T)\}$ is the zero-mean, unit-variance fMRI BOLD time series for region $i$.

\subsection*{Dynamic functional connectivity}
We also constructed each participant's dynamic functional connectivity (dFC) matrix, $\mathcal{W}=\{\mathbf{W}(1),\ldots,\mathbf{W}(T-L+1)\}$, where $\mathbf{W}(t)=[W_{ij}(t)]$ is the estimated connectivity matrix for the time interval beginning at $t$. This entailed first dividing the regional BOLD time series into overlapping windows of approximately 100 seconds in length. With a sampling frequency of $f=\frac{1}{0.645}$ Hz, this translated to a window of $L=156$ time points. The decision to select a window of 100 s was made so that the window was long enough to capture a full cycle of the slowest frequency component. As the high-pass cutoff for the BOLD time series was 0.01 Hz, the shortest possible window was 100 s long \citep{zalesky2015towards, leonardi2015spurious}. For each window, we then calculated the cross-correlation matrix using only the observations within that window. The cross-correlation was calculated after exponentially discounting the fluctuations of more distant time points so that the correlation coefficients weighed recent events more heavily.

In slightly more detail, we define a discounting function for each window:

\begin{equation}
w(\tau) = w_0 e^{(\tau - L)/\theta}
\end{equation}

where $\tau = 1, \ldots, L$, $w_0 = (1 - e^{-1/\theta})/(1 - e^{-L/\theta})$, and $\theta = L/3$.

Based on this weighting, we calculated for each window:

\begin{equation}
\bar{z}_i (t) = \sum_{\tau = 1}^{L} w(\tau) z_i (t - L + \tau)
\end{equation}

\begin{equation}
\sigma_i (t) = \Big[ \sum_{\tau = 1}^{L} w(\tau) (z_i (t - N + \tau) - \bar{z}_i (t)) \Big]^{1/2}
\end{equation}

\begin{equation}
\sigma_{ij} (t) =  \Big[ \sum_{\tau = 1} ^ {L} w(\tau) (z_i (t - N + \tau) - \bar{z}_i (t)) \cdot (z_j (t - N + \tau) - \bar{z}_j (t)) \Big]^{1/2}
\end{equation}

The estimated functional connectivity at time $t$ was then calculated as $W_{ij}(t) = \sigma_{ij}(t)/(\sigma_i(t)\sigma_j(t))$.

From $\mathcal{W}$, we can estimate the variability in the fluctuations of the functional connection between nodes $i$ and $j$ over time as:
\begin{equation}
s_{ij}=\sqrt{\frac{1}{T-L}\sum_{t=1}^{T-L+1}(W_{ij}(t) - m_{ij})}
\end{equation}
where $m_{ij}=\frac{1}{T-L+1}\sum_{t=1}^{T-L+1}W_{ij}(t)$ is the mean dynamic functional connectivity over time.

\subsection*{Conditional dynamic functional connectivity}
\begin{figure}[ht]
\centering
\includegraphics[width=\linewidth]{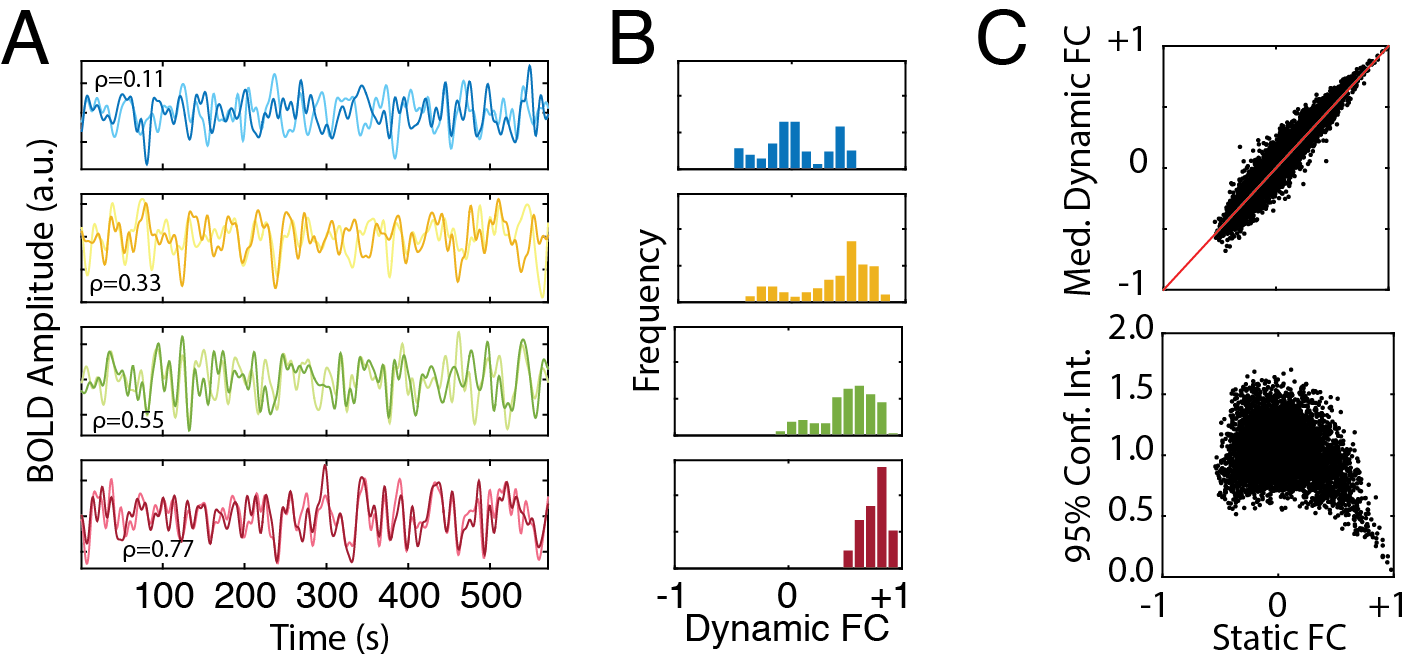}
\caption{Functional connectivity at long time scales constrains the range over which functional connectivity is expected to vary at shorter time scales (A) Four pairs of real BOLD time series whose static connectivity varies from $\rho=0.11$ (blue), $\rho=0.33$ (yellow), $\rho=0.55$ (green), and $\rho=0.77$ (red). (B) We then sub-divided the full time series into 730 windows containing 156 time points each (the same parameters used for generating dynamic FC matrices), estimated dynamic FC for each window, and then plotted the histogram of dynamic FC magnitude. (C) In general, the distribution is centered around the magnitude of the static connection and also becomes tighter and more skewed as the static connection increases in magnitude. The top panel shows the magnitude of static connectivity on the x-axis and the median (50th percentile) of dynamic connectivity on the y-axis. The two curves are correlated to a value of 0.95. The bottom panel shows the magnitude of static connectivity plotted against the range of the 95\% confidence interval for the distribution of dynamic fluctuations (i.e. we expect that 95\% of all dynamic functional connection magnitudes will fall within this range).}
\label{fig:fig1}
\end{figure}

It is important to recognize that the strength of static functional connectivity observed over a long time period imposes constraints on the extent to which dynamic functional connectivity is expected to fluctuate over shorter time scales within that same period \citep{thompson2015mean}. We can think of these constraints in the following way. Suppose we have two BOLD time series: $\mathbf{x}=\{x(1) \ldots x(T) \}$ and $\mathbf{y}=\{y(1) \ldots y(T) \}$ that are correlated with magnitude $\rho_{xy}$. Now, suppose we were to divide both $\mathbf{x}$ and $\mathbf{y}$ into sub-intervals and calculate each sub-interval's correlation coefficient. It turns out that the magnitude $\rho_{xy}$ plays an important role in constraining the distribution of sub-interval correlation coefficients (Figure \ref{fig:fig1}). Because of these constraints, we can make predictions about the expected range of dynamic fluctuations given a connection's sFC. Observing a strong static correlation coefficient makes it highly unlikely, if not physically impossible, for the corresponding dFC to fluctuate over a wide range. Instead, observing a weak static correlation coefficient is compatible with much more pronounced fluctuations, including temporary expression of strong positive or strong negative dFC. Hence, the interpretation of fluctuations in dFC will vary depending on whether the corresponding sFC is weak or strong.

A corollary of this approach is that fluctuations in dynamic functional connectivity may be unsurprising, in a statistical sense, unless they violate expectations conditioned by the level of sFC. We argue that the interpretation of dFC should take into account whether observed features of dFC are expected or unexpected. Doing so allows one to uncover fluctuations in dynamic functional connectivity between regions that are unexpectedly strong or weak given those regions' static connectivity. To characterize such fluctuations, we propose a measure that we term \textit{conditional dynamic functional connectivity} (or cdFC), which is defined as the probability of observing a dynamic functional connection of a particular magnitude given the magnitude of static connectivity for the same pair of brain regions.

To estimate the cdFC for any dynamic functional connection, we first characterize the distribution of expected dynamic fluctuations given a connection's static connectivity. There are many ways to approximate this distribution; the approach we use is to generate surrogate BOLD time series using the amplitude adjusted phase randomization procedure. Briefly, this procedure works by generating a normally-distributed surrogate time series that is rank-matched to the original time series. The surrogate time series is then transformed into the frequency domain via a discrete Fourier transform. Random phase (between $[0, 2\pi]$) is then added to each frequency bin. If this procedure is being performed on multiple time series, the random phase is added uniformly across all channels to the same frequency bin. Next, the data are transformed back into time series via an inverse Fourier transform. Finally, the original time series amplitudes are rank-matched to the phase-randomized time series. The resulting surrogate time series preserve the same amplitude distribution of the original data and approximate the same power spectrum.

\begin{figure}[ht]
\centering
\includegraphics[width=\linewidth]{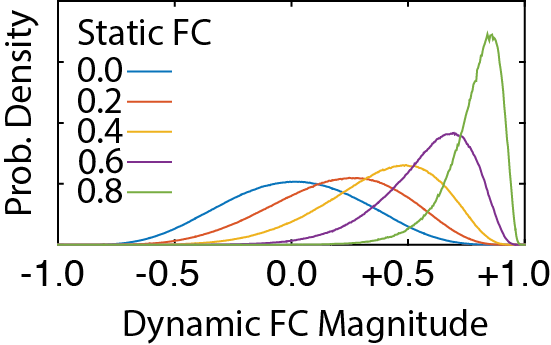}
\caption{Expected dynamic functional connection distributions for a range of static connectivity magnitudes. Each color represents a different magnitude static connection: 0.0 (blue), 0.2 (orange), 0.4 (yellow), 0.6 (purple), and 0.8 (green). These curves were estimated by amassing the dynamic functional connections estimated from surrogate BOLD time series.}
\label{fig:fig2}
\end{figure}

For each realization of surrogate data, we generate a dynamic functional connectivity matrix, from which we collect observations of dynamic fluctuations. Repeating this procedure many times allows us to approximate the distribution of dynamic fluctuations for each functional connection (Figure \ref{fig:fig2}). We return to the original data and assign each dynamic functional connection a percentile based on where that dynamic fluctuation falls with respect to the distribution of expected fluctuations.

It should be noted that there are alternative methods for estimating this null distribution. For example, instead of estimating this distribution using phase-randomized surrogates, one could follow the approach of \cite{zalesky2014time} and estimate the parameters of a bi-variate time series model (e.g. autoregressive moving average - ARMA) for a dynamic functional connection. The ARMA model and its parameters could then be used to generate surrogate time series and dynamic functional connectivity matrices, and from these matrices one could approximate the distribution of expected fluctuations.

\subsection*{Characterizing conditional dynamic functional connectivity}
\begin{figure}[ht]
\centering
\includegraphics[width=\linewidth]{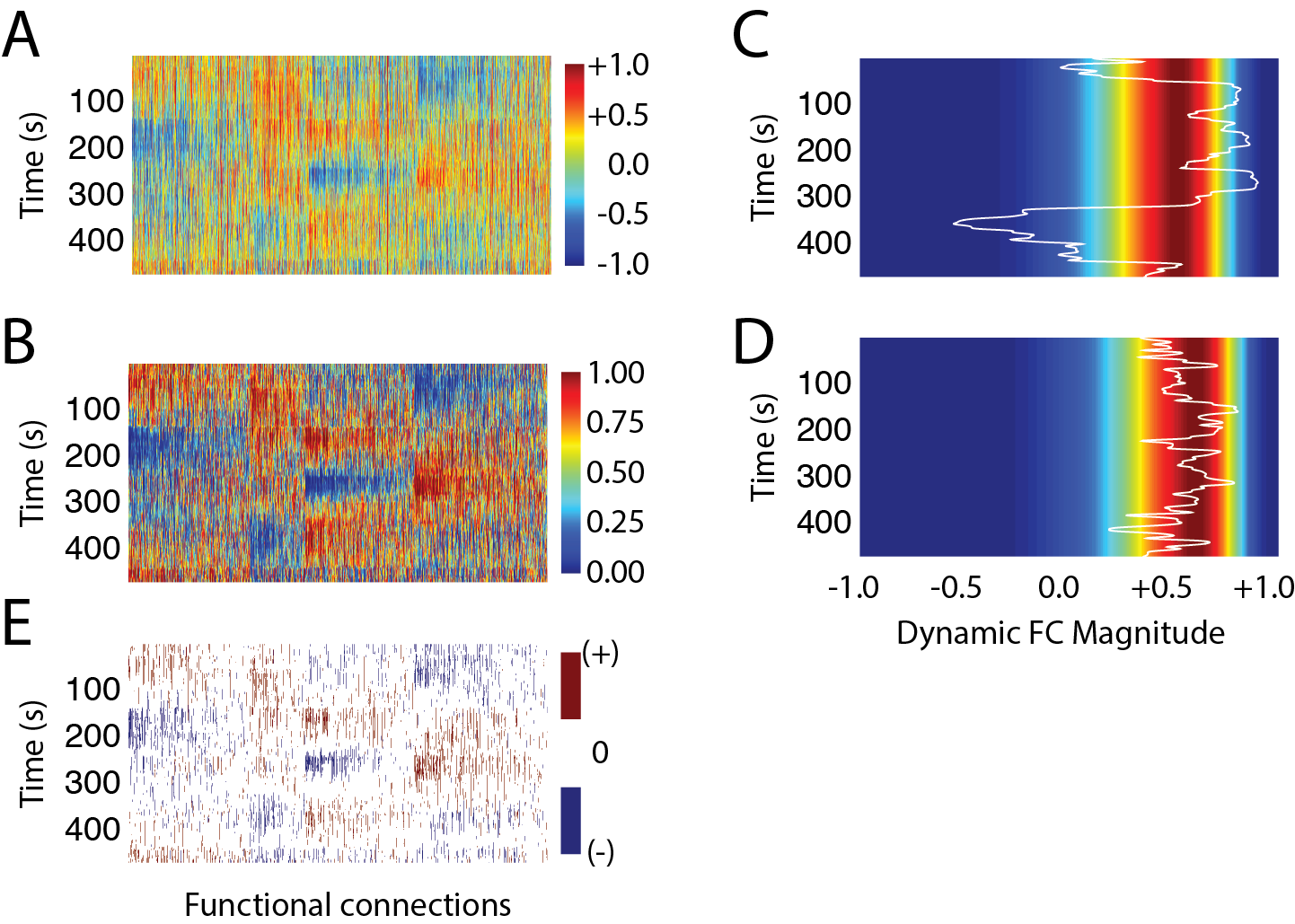}
\caption{Transformation of dynamic FC to conditional dynamic FC. (A) Observed dynamic FC matrix. (B) Conditional dynamic FC matrix. (C) Example FC time series with many fluctuations beyond the the 95\% confidence intervals. (D) Example FC time series where all fluctuations fall within the 95\% confidence interval. The time series shown in (C) and (D) were selected because they had similar values of static connectivity. (E) Thresholded matrix, highlighting fluctuations that are stronger than expected (shown in red) and fluctuations that are weaker than expected (shown in blue). The x-axes of panels A, B, and E represent functional connection indices. Each column in those matrices represents the temporal evolution of one of the $n(n - 1)/2=6441$ functional connections.}
\label{fig:fig3}
\end{figure}

We made several measurements based on the conditional dynamic functional connectivity matrices. First, we applied a threshold to each matrix, retaining only those connections that were unexpectedly strong or weak (greater than the 97.5th percentile or less than the 2.5th percentile; see Figure \ref{fig:fig3}). From these thresholded matrices, we calculated the number of unexpectedly strong/weak connections at each time point, $E^+(t)$ and $E^-(t)$, respectively, which we refer to as \textit{excursions}. We also counted the total number of excursions of both types at each time $t$ as $E(t) = E^+(t) + E^-(t)$. To contextualize the number of excursions at any time point, we compared the observed counts to those obtained from an additional sample of 1000 surrogate datasets generated as described before. This allowed us to assign each excursion count score to a percentile and to focus on time points at which the number of excursions was greater than expected (greater than or equal to the 97.5th percentile).This process of contextualizing dynamic functional connections can be seen from a statistical point of view as an assessment of whether there is enough evidence to reject the null hypothesis that a dynamic functional connection is neither significantly stronger or weaker than its corresponding static connection.

It should be noted that the characterization of a dynamic functional connections as stronger or weaker than expected is always made with respect to a  connection's static weight and never in an absolute sense. In other words, ``stronger'' refers to ``more positive than expected'' and ``weaker'' means ``more negative than expected.'' This can lead to confusion when dealing with static connections whose weights are $<0$. In such cases, weaker-than-expected actually means a more negative correlation (i.e. close to $-1$) while stronger-than-expected means a more positive correlation (i.e. closer to $+1$). As a consequence of adopting this convention a stronger-than-expected dynamic functional connection may actually approach a value of 0.

An alternative, and admittedly simpler, method for contextualizing excursion counts would be to randomly permute the order of time points, repeat this process many times, and compare the observed excursion counts to the distribution generated from the permutations. We believe that this approach, though simpler, is not appropriate, because dynamic functional networks are not independent of one another due to overlapping time windows. For example, the observations used to construct the dynamic networks at time $t$ and $t + 1$ have $L - 1$ points in common, where $L$ is the window length. Random reorderings of the excursion count time series would likely violate this interdependency. This procedure can also be viewed from a statistical point of view. In short, we want to test the null hypothesis that a dynamic functional connection is either significantly stronger or weaker than its corresponding static connection.

We also calculated the global excursion matrix, $\mathbf{X}$, whose element $x_{ij}$ was equal to the number of times, across all subjects, that the connection $x_{ij}$ participated in an excursion. As before, we were able to divide $\mathbf{X}$ into $\mathbf{X}^+$ and $\mathbf{X}^-$, which count the number of times a connection was stronger or weaker than expected, respectively.

\subsection*{Community detection}
We used the dynamic functional connectivity matrices (in their raw form) in conjunction with modularity maximization \citep{newman2004finding} to identify functional communities \citep{sporns2015modular} and to assess their quality. This process entailed maximizing the \textit{modularity} quality function for the dynamic functional connectivity matrix at time $t$:

\begin{equation}
Q(t)= \sum_{ij} B_{ij}(t) \delta(g_i(t),g_j(t))
\end{equation}

Here, $\mathbf{B}(t)=[B_{ij}(t)]$ is the so-called modularity matrix, and is equal to $\mathbf{B}(t) = \mathbf{W}(t) - \mathbf{P}(t)$, where $\mathbf{P}(t)=[P_{ij}(t)]$ is the expected weight of the functional connection between brain regions $i$ and $j$ at time $t$. The function $\delta(g_i(t),g_j(t))$ is the Kronecker delta and is equal to unity when the community assignments of regions $i$ and $j$ at time $t$, $g_i(t)$ and $g_j(t)$ respectively, are the same. Otherwise the delta function is equal to zero. Traditionally, the definition of $\mathbf{P}$ is left up to the user. Recent work has shown that $\mathbf{P}$ must be chosen carefully when one applies modularity maximization to correlation matrices. One possible choice is to set $\mathbf{P}=\mathbf{I}$. where $\mathbf{I}$ is the identity matrix \citep{macmahon2013community, bazzi2014community}. This definition corresponds to a null model where time series are uncorrelated with one another. Thus, community detection amounts to placing as many positive correlation coefficients within each community as possible.To normalize $Q$ so that it is bounded between 0 and 1, we scale $Q(t)$ by $1/2m$, where $2m = \sum_{ij} |W_{ij}|$.

%%%%%% new section %%%%%%%%%

It is worth noting that the modularity function used here differs from most standard approaches for dealing with signed networks. Traditionally, signed networks are sub-divided into two separate networks: one containing just positive connections:

\[
    W_{ij}^+= 
\begin{cases}
    W_{ij} ,& \text{if } W_{ij} > 0\\
    0,              & \text{otherwise}
\end{cases}
\]

and 

\[
    W_{ij}^-= 
\begin{cases}
    -W_{ij} ,& \text{if } W_{ij} < 0\\
    0,              & \text{otherwise}
\end{cases}
\]

Modularity functions are then defined for both the positive and negative components as: $Q^\pm = \sum_{ij} [W_{ij}^\pm - P_{ij}^\pm ]\delta (g_i,g_j)$.  The choice of how to combine the two components to obtain the total modularity is left up to the user, though a number of weighting schemes have been proposed \citep{gomez2009analysis, rubinov2011weight}. In general, the total modularity is of the form: $Q= c^+ Q^+ - c^- Q^-$, where $c^\pm$ are constants. Thus, modularity maximization for signed networks can be viewed as an attempt to maximize the modularity of positive connections while penalizing negative connections when they fall within communities.

The approach we use here does not explicitly distinguish between positive and negative connections. However, because we essentially treat $W_{ij} = B_{ij}$, it means that only positive functional connections can contribute to $Q$ and negative connections, when they fall within communities, serve to decrease the total modularity.

%%%%%%%%%%%%%%%%%%%%%%

We adopt this approach here, and use a Louvain-like algorithm \citep{blondel2008fast, jutla2011generalized} to maximize modularity 100 times for each dynamic functional connectivity matrix. The output of the Louvain algorithm is a set of community assignments, which specifies a partition of the brain, along with a score ranking the quality of this partition. The number and size of communities, in general, partially determined by the total density of the connections \citep{fortunato2007resolution}, but can be varied by including a resolution parameter in the modularity expression that tunes the size and number of detected communities \citep{reichardt2006statistical}. While this type of multi-resolution modularity maximization has been used to show that brain networks exhibit interesting modular structure at multiple organizational scales \citep{betzel2013multi}, here we focus on a single-scale estimate of community structure.

At the particular scale we investigate, the Louvain algorithm's output typically varies from run to run. Most analyses regard such variability as problematic and attempt to resolve it by constructing, from the varied outputs, a \textit{consensus partition} that represents the average community structure \citep{lancichinetti2012consensus}. Here, we embrace this variability and use it to characterize the ``modularity landscape'' for a given dynamic functional connectivity matrix. The modularity landscape refers to the space of all possible community divisions (which is far too large to enumerate in its entirety). Each point in this landscape can be assigned a fitness score equal to $Q$. We can think of the Louvain algorithm as moving through this landscape, looking for solutions of increasingly greater fitness (highly-modular partitions). If the landscape features a single highly-fit partition (meaning that this partition is much more modular than other nearby partitions), we expect the Louvain algorithm to successfully negotiate the landscape and, more times than not, return as its output a partition that approximates the optimal partition. However, if the landscape features multiple near-optimal partitions we expect the Louvain algorithm to have difficulty arriving at the same solution each run. We refer to this type of landscape as a \textit{degenerate} or \textit{near-degenerate} modularity landscape, indicating the presence of multiple near-optimal solutions \citep{good2010performance}.

With this intuition in mind, we calculated the similarity of all pairs of detected partitions at time $t$ (a total of $100(100 - 1)/2 = 4950$ pairs) using the z-score of the Rand index \citep{traud2011comparing} and subsequently averaged over all pairs, thus obtaining a single score, $Z(t)$, that quantified the average pairwise similarity of partitions at time $t$. If $Z(t)$ was large, it indicated the the Louvain algorithm consistently returned similar partitions. On the other hand, if $Z(t)$ was small, then the partitions varied from run to run. We interpreted the magnitude of $Z(t)$ as an indication of the level of degeneracy in the modularity landscape, with high values indicating low levels of degneracy. It should be noted that the Louvain algorithm uses a greedy heuristic to maximize modularity and very likely samples near-optimal partitions in a somewhat biased manner.

\section*{Results}

\subsection*{The resting state alternates between periods of unexpectedly strong/weak connectivity and high/low modularity}

\begin{figure}[ht]
\centering
\includegraphics[width=\linewidth]{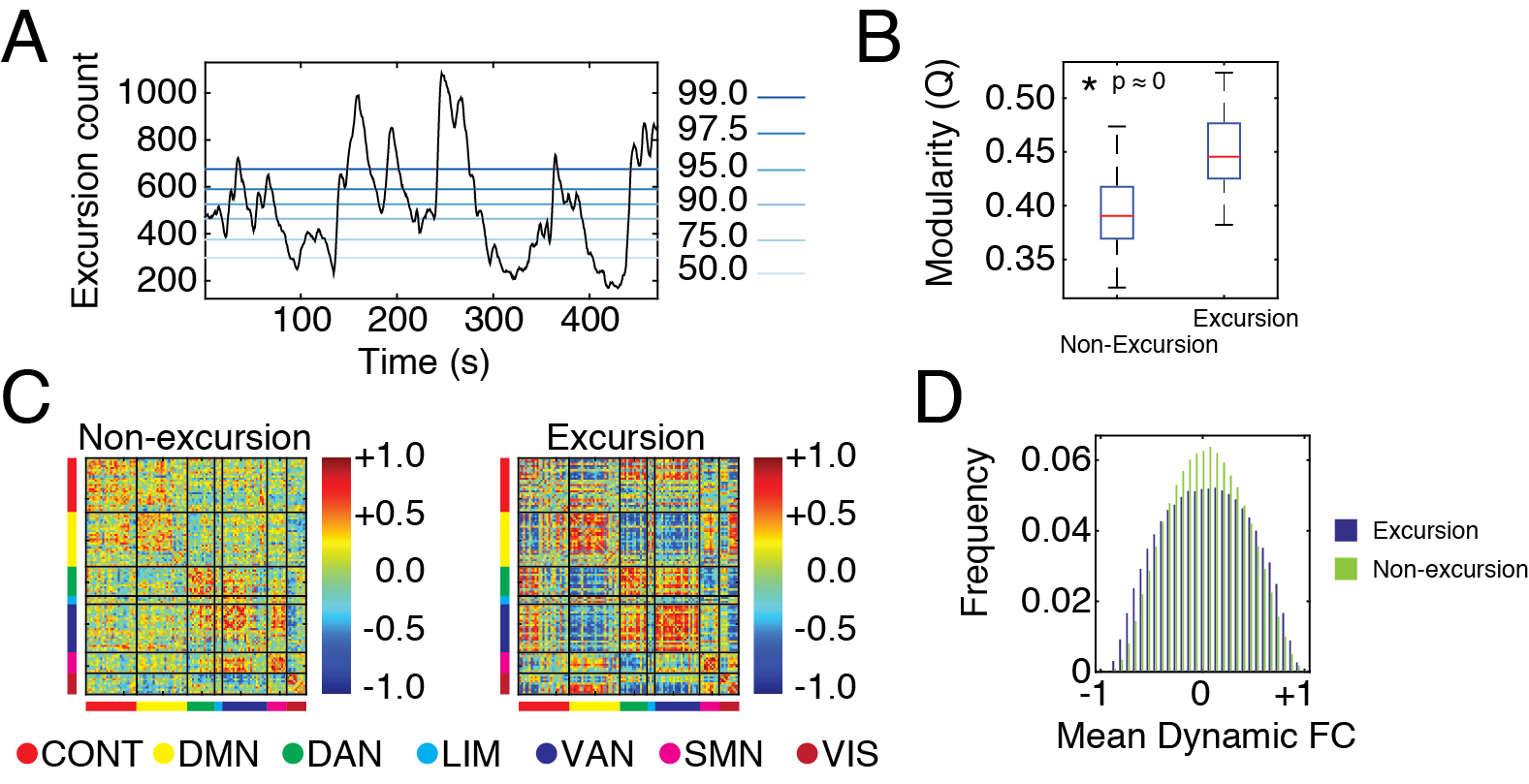}
\caption{Identification and characterization of mass excursions. (A) Raw excursion count (shown in black), $E(t)$, along with null percentiles ranging from 50\% to 99\%. (B) The modularity during mass excursions was, on average, greater than the modularity during non-excursions, an effect that was likely driven by the increased number of strong functional connections. (C) We show functional network topologies for a non-excursion and an excursion. The module labels come from the Yeo2011 intrinsic connectivity network partition, which divides the brain into seven functional systems (and seventeen sub-systems not shown here). The systems correspond to control (CONT), default mode (DMN), dorsal attention (DAN), limbic (LIM), ventral attention/saliency (VAN), somatomotor (SMN), and visual (VIS) networks. (D) We compare the edge connection weight distribution for non-excursions with the distribution during excursions. During excursions the distribution is broader and includes values closer to $\pm1$.}
\label{fig:fig4}
\end{figure}

We estimated conditional dynamic functional connectivity for $N=80$ individuals and identified, for each connection, the time points at which it was unexpectedly strong or weak. Averaged across participants, this procedure classified roughly $6.7\pm0.8\%$ of all dynamic functional connections as excursions. We calculated the total number of excursions at each time point, $E(t)$, and compared this number to a null model to identify instants where the total number of excursions was greater than chance (Figure \ref{fig:fig4}A). On average, this procedure identified $143\pm75$ time points (of 730 possible) as \textit{mass excursions}. The periods during which many connections were collectively unexpectedly strong/weak produce functional brain networks that corresponded to functional networks with topologies that were, on average, more modular than those encountered at time points with fewer excursions (t-test, df = 728, 74 of 80 participants had p-value less than $p=0.01$; of those participants the median p-value was $p \approx 10^{-19}$). This effect was, in part, driven by the distribution of functional connections during mass excursions, which favor extreme correlations (values near $\pm1$) and make it possible for the network to achieve highly modular configurations.

\subsection*{Event count is correlated with low-degeneracy of modularity landscapes}
\begin{figure}[ht]
\centering
\includegraphics[width=\linewidth]{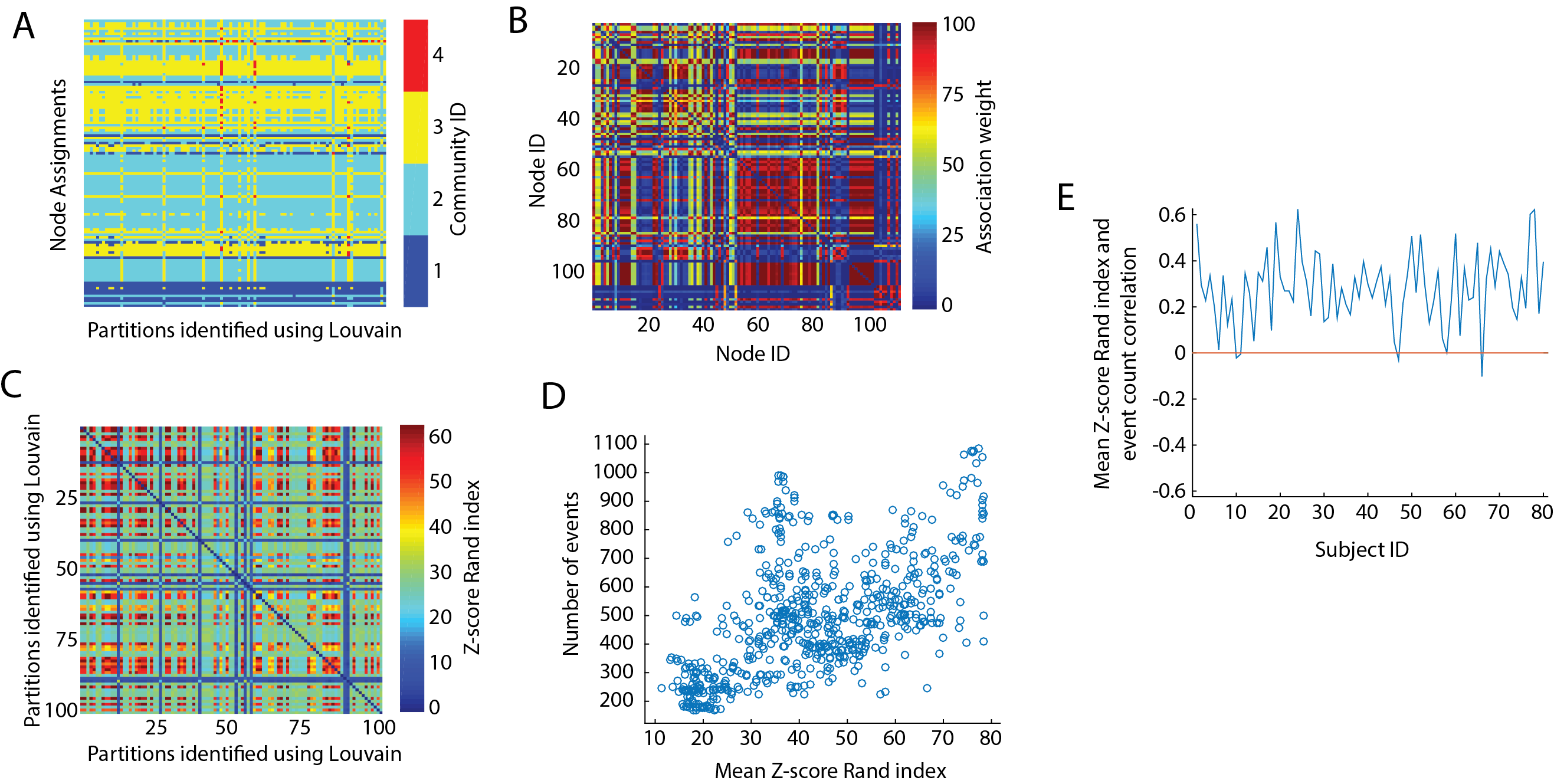}
\caption{Partition degeneracy and correlation with excursion count. (A) As an example, we show the set of 100 partitions obtained by maximizing modularity for a single instance of dFC (169 events). (B) The association matrix for this set of partitions. Each element in the association matrix counts the number of times that two nodes were assigned to the same partition. The node order is the same as in Figure 4. (C) Matrix of z-score Rand indices for all pairs of the 100 partitions shown in panel (A). (D) We show the scatterplot of mean z-score Rand index for each time point plotted against the event count, both calculated at the same time point, for one representative participant. (E)  Correlation coefficients for z-score Rand index and event count plotted across participants. Note that of the 80 participants, only four exhibited negative correlations.}
\label{fig:fig5}
\end{figure}

In addition to calculating the average modularity of dynamic functional connectivity networks over time, we also calculated the average similarity of partitions generated by the Louvain algorithm. We found that, across all participants, partition similarity was positively correlated with event count (median correlation $r = 0.19$, interquartile range of $[0.09, 0.32]$) (Figure \ref{fig:fig5}). The similarity of partitions is related to the degeneracy of the modularity landscape. In other words, if there are many highly-modular, yet dissimilar, partitions, then we would expect the Louvain algorithm to return a diverse set of partitions; sometimes the algorithm would converge to one partition while other times it would converge to another. However, if the maximum modularity corresponds to only a small number of similar partitions, we expect the algorithm to find and converge to these few solutions. Thus, we can interpret the similarity measure as indicating the level of the degeneracy in the landscape of partitions.

It is important to note that the use of Pearson's correlation or other similar measures for estimating functional connectivity may make it easier for many connections to simultaneously become stronger or weaker than expected. This is a consequence of \textit{transitive correlations} and the fact that functional connections are not independent of one another \citep{zalesky2012use}. In other words, if a small set of regions becomes more strongly or weakly correlated with one another, then other regions with which any of these regions are also correlated will likely experience a similar effect. This could potentially cause us to over-estimate the number of independent excursions at any time point and hence the size of mass excursions.

Another important consideration is that excursion count might be driven by extraneous (or otherwise unwanted) factors, such as head motion \citep{power2012spurious, power2014methods}. To this end, we compared excursion count to the frame-wise displacement time series, which we processed in precisely the same way as the BOLD time series, so that the motion estimate at each time point was a weighted average of displacements that occurred within a window of time. We find no consistent group-level relationship between motion estimates and excursion counts when we aggregate these variables across all participants (Pearson's correlation coefficients of $r = 0.044$ and $r=0.041$ for frame-wise displacement measured as L1 and L2 norms from a reference volume). At the level of individual participants, however, we found cases where event count was significantly correlated with the frame-wise displacement time series, though this correlation was neither consistently positive nor was it consistently negative (21/80 positively and 23/80 negatively correlated at a significance level of $p=0.01$, Bonferroni-corrected; mean$\pm$standard deviation Pearson's correlations of $r=-0.016\pm0.223$ and $r=-0.025\pm0.223$ for L1 and L2 norm). Thus, while the group-level and participant-level analyses suggest that motion is not systematically related to excursion counts, we do observe that some participants show significant correlations (both positive and negative), and hence motion may act as a potential confound in a subset of participants. Interestingly, the number of detected communities is also significantly and negatively correlated with excursion in most individual participants (median correlation $r= -0.20$, interquartile range of $[-0.32, 0.06]$).

\subsection*{Default mode network dissociates during mass excursions}
\begin{figure}[ht]
\centering
\includegraphics[width=\linewidth]{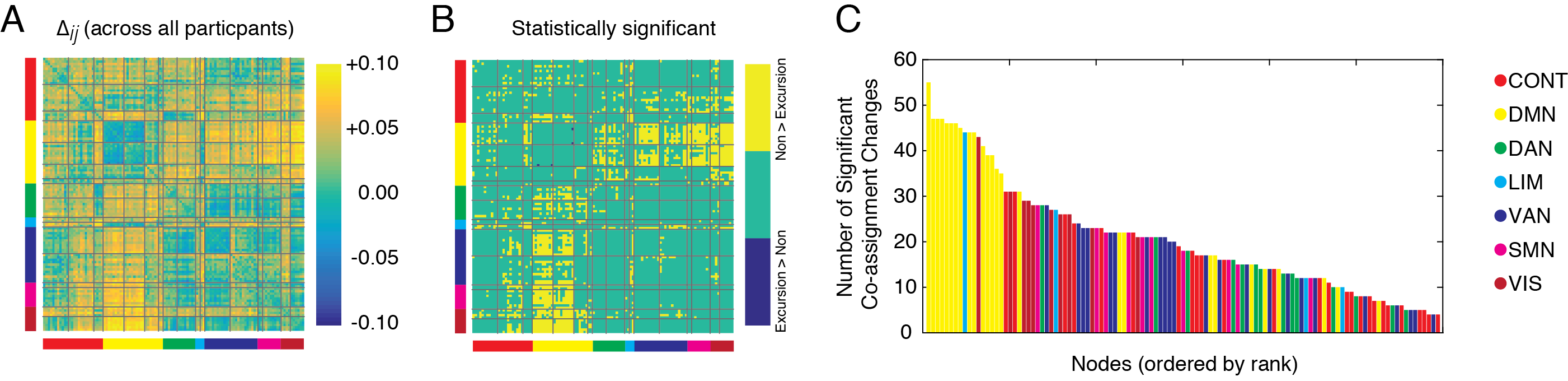}
\caption{Default mode network dissociates during mass-excursions. A) Mean difference in association matrices constructed during non-mass excursions and excursions. B) Node pairs that were statistically more or less likely to be observed in the same community during non-excursions compared to excursions. The network labels are taken from \cite{yeo2011organization}, with labels corresponding to networks for control (CONT), default mode (DMN), dorsal attention (DAN), limbic (LIM), ventral attention (VAN), somatomotor network (SMN), and visual (VIS). C) We sum the rows of the matrix in panel B) to identify regions that that consistently change their community co-assignments with other regions. The regions that change the most are predominantly regions associated with DMN.}
\label{fig:fig6}
\end{figure}

IIn addition to quantifying the degeneracy of modular architectures, we also examined whether community structure was consistently different during mass excursions compared to non-mass excursions. To detect such differences, we first divided each participant's dynamic functional connectivity networks into two classes: one class corresponding to mass-excursion events and another class corresponding to all other instants. The previous community detection procedure yielded 100 estimates of the community structure for each dynamic functional connectivity network, which we aggregated according to class. From these data and for each class we generated an association matrix, $\mathbf{T}$, which was a square, $n \times n$ matrix whose elements $T_{ij}$ represented the probability, across all community estimates assigned to that class, that nodes $i$ and $j$ were assigned to the same community. Association matrices, unlike ``hard'' partitions that assign nodes to one class or another, give a quasi-continuous estimate of whether two nodes appeared in a community together. Next we calculated the element-wise difference in association matrices, subtracting the mass excursion association matrix from the non-mass excursion matrix. The elements of the resulting matrix, $\Delta_{ij}$, quantified how much more likely it was for two nodes to be assigned to the same community during non-mass excursions compared to mass excursions (Figure \ref{fig:fig6}A). We performed this analysis for all participants and asked, for each node pair $\{ i,j \}$ whether $\Delta_{ij}$ was consistently greater or less than zero across all subjects (mass-univariate t-test with statistical threshold of $p<0.05$, FDR-corrected) (Figure \ref{fig:fig6}B).

Interestingly, the regions that consistently changed their community co-assignments were, disproportionately so, regions associated with default mode network (DMN). In particular, during non-mass excursions DMN regions are grouped with regions that form parts of the dorsal attention, ventral attention, somatomotor, visual, and control systems, whereas during mass excursions these same DMN regions dissociate from these systems. The regions that changed their community co-assignments most frequently were the bilateral posterior cingulate, dorsal and medial prefrontal cortex along with the inferior parietal lobule, all subcomponents of the DMN  (Figure \ref{fig:fig6}C).

\subsection*{Excursions occur most frequently for connections within visual and somatomotor networks}

\begin{figure}[ht]
\centering
\includegraphics[width=\linewidth]{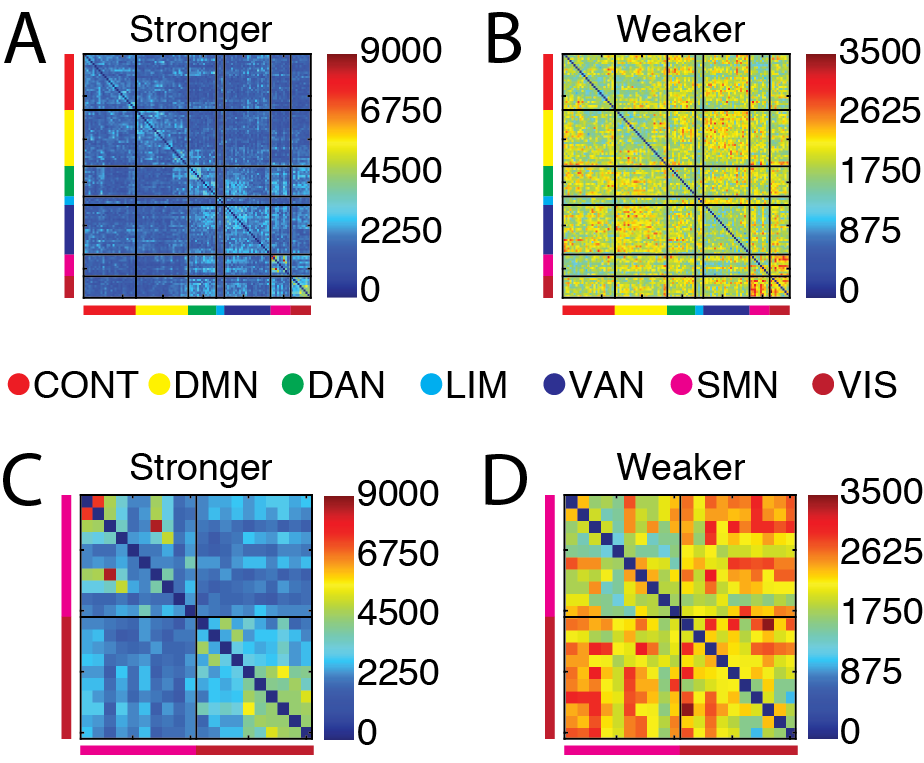}
\caption{Connection-wise excursion count matrices. (A) The number of times across all participants that a functional connection achieved a greater-than-expected level of functional connectivity, $\mathbf{X}^+$. (B) A matrix similar to that shown in panel (A), but shows connections that achieved weaker-than-expected (more anti-correlated than expected) levels of functional connectivity, $\mathbf{X}^-$. The colored bars along the x and y axes indicate the intrinsic connectivity network to which each region was assigned. The network labels are taken from \cite{yeo2011organization}, with labels corresponding to networks for control (CONT), default mode (DMN), dorsal attention (DAN), limbic (LIM), ventral attention (VAN), somatomotor network (SMN), and visual (VIS). (C) and (D) Enlargement of panels (A) and (B) highlighting the connections within and between visual and somatomotor networks.}
\label{fig:fig7}
\end{figure}

In order to identify whether excursions were driven by any particular functional connections or whether certain connections participated disproportionately in a greater number of excursions, we counted the number of times that a connection, $\{i,j\}$ was stronger than expected, $x^+_{ij}$. We did the same for weaker-than-expected connections, $x^-_{ij}$. We found that, across all participants, all connections participated in at least one excursion. Interestingly, however, a small number of connections participated in disproportionately many (Figure \ref{fig:fig7}). Most conspicuous were homotopic connections between left and right central sulcus, primary somatosensory area, striate, and superior extra-striate cortex. Similarly, the connections that were most frequently weaker than expected were dominated by those involving visual and somatomotor networks (see Figure \ref{fig:fig7}C-D for more detail of the connections between visual and motor networks).

\begin{figure}[ht]
\centering
\includegraphics[width=\linewidth]{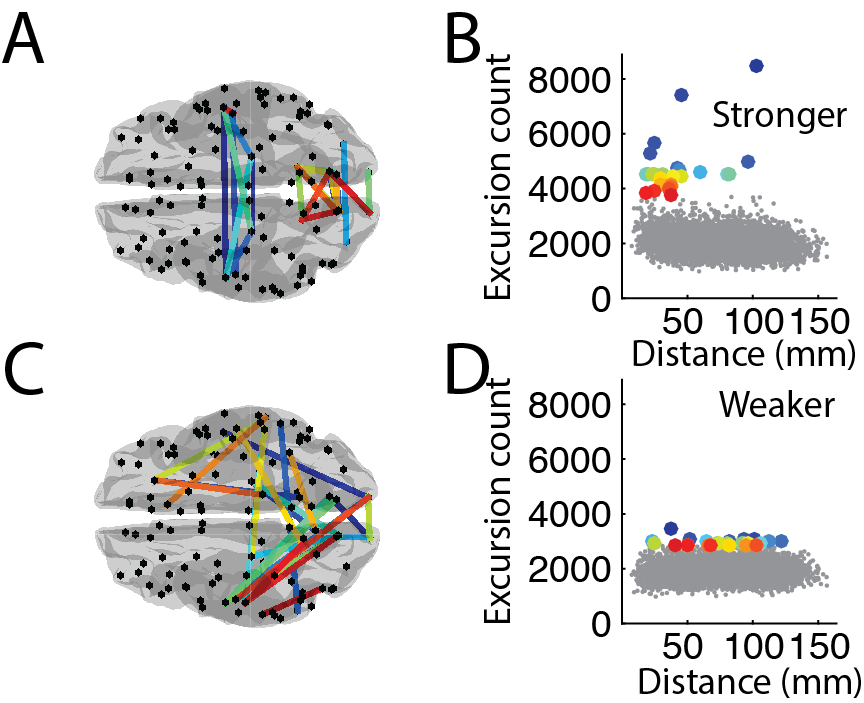}
\caption{Anatomical distribution of stronger-/weaker-than-expected functional connections. In all panels A-B we focus on the 25 connections that appear most frequently as stronger-than-expected. In panels C-D we focus on the 25 connections that are consistently weaker than expected. The connections are color-coded in both cases so that they can be identified in the anatomical plots as well as the distance vs event count plots.}
\label{fig:fig8}
\end{figure}

Portions of the visual (VIS) and somatomotor (SMN) networks are situated near the midline dividing the left and right cerebral hemispheres. One possibility is that these connections appear disproportionately often due to the short Euclidean distance between the brain regions that make up those networks (i.e. a bias towards shorter functional connections). While it is difficult to completely discount this possibility (and some of the most consistently strong/weak connections are short-range), many are long-range, especially those involving their homotopic partners (see Figure \ref{fig:fig8}E), suggesting that whether or not a connection participates in an excursion is not a mere consequence of its length. To verify that distance is not the principal driver of excursion counts, we calculated the Spearman rank-correlation of $x^-_{ij}$, $x^+_{ij}$, and $x^-_{ij} + x^+_{ij}$ with Euclidean distance and found, $r_- = 0.14$, $r_+=0.035$, and $r_{-,+} = -0.092$. Each of these correlations were significant at $p<0.005$, due mainly to the fact that the correlations included many observations ($113\times112 / 2$ points); the actual magnitudes of the correlations were small in all cases.

We also wished to determine what was the influence, if any, of motion on this result. To this end, we divided our population of $N=80$ participants into two groups based on the magnitude with which their excursion count time series was correlated with framewise displacement. The cutoff correlation for the ``low motion correlation'' group was $|r_{motion,excursion count}| < 0.15$. This resulted in a division of our participant pool into 38 low- and 42 high-motion correlation groups. For each group, we generated, separately, estimates of $x^-_{ij}$, $x^+_{ij}$. We then asked whether the elements of these matrices from the low-motion correlation group were similar to those generated from the full cohort. We found a strong correspondence in both cases, with matrix-wise correlations of $r^+=0.85$ and $r^-=0.75$ (Figure \ref{fig:sifig4}).

\section*{Discussion}

In this report we focus on an aspect of dynamic functional connectivity that has so far been left relatively unexplored. We show that, when measured using Pearson's correlation coefficient, certain aspects of dynamic functional connectivity, namely the temporal mean and the range of dynamic fluctuations, can be partially predicted from the static connectivity alone. Our analyses indicate that this relationship is, in part, a mathematical consequence of time series dynamics and may therefore not implicate any underlying neurobiological process that actively drives dFC fluctuations. To account for this statistical relationship, we propose a measure that highlights fluctuations in functional connectivity that are unexpected given the underlying static connectivity. We use this method to identify time points when specific connections are unexpectedly strong or weak, and find that these connections tend to cluster temporally, resulting in mass excursions. We find that during these events, functional brain networks adopt highly modular topologies compared to other time periods. Furthermore, we show that such events tend to involve a disproportionately large number of connections associated with visual and somatomotor systems compared to higher-level association networks. We go on to show that, across participants, these events are not systematically related to participant head motion and include many long-range connections, suggesting that they cannot be explained as spatial artifacts.

Computational models have suggested that functional connectivity and its associated network characteristics are time-dependent and vary over multiple time scales \citep{honey2007network, hansen2015functional}. Empirical functional connectivity analyses are beginning to move beyond static or time-invariant studies, focusing increasingly on the ongoing temporal dynamics of brain connectivity \citep{hutchison2013dynamic, calhoun2014chronnectome, kopell2014beyond}. Biomarkers based on dynamic functional connectivity appear to track behaviorally relevant variables \citep{elton2015task}, including daydreaming \citep{kucyi2014dynamic} and state of arousal \citep{allen2012tracking}, cognitive state \citep{betzel2012synchronization, hutchison2013resting, tagliazucchi2014enhanced}, consciousness \citep{barttfeld2015signature}, and developmental level \citep{hutchison2015tracking}. They may also prove useful in differentiating between diseased and healthy populations \citep{damaraju2014dynamic, rashid2014dynamic, miller2014higher}. Despite these initial successes, there are a number of methodological and interpretational issues related to estimating and interpreting dynamic functional connectivity. For example, while the \textit{sliding window} approach used here is most common, other methods for estimating dynamic functional connectivity exist, such as component \citep{calhoun2012multisubject} and model-based analyses \cite{lindquist2014evaluating}. The advantage of one method over the other is not immediately clear. In terms of how we interpret dynamic functional connectivity, it remains unclear whether fluctuations in fMRI connectivity strength over time have a distinct neurobiological basis, i.e. are the expression of fluctuations in  the underlying pattern of neural activity, though a number of studies using combined EEG-fMRI acquisition have reported electrophysiological correlates of fMRI fluctuations \citep{scheeringa2012eeg, tagliazucchi2012dynamic, chang2013eeg} . An alternative interpretation is that these fluctuations merely reflect correlated white noise \citep{hlinka2015danger}, or can be accounted for by the configuration of \textit{static functional connections} \citep{thompson2015mean}. The latter effect was the starting point for the analysis conducted in this paper.

\subsection*{Mass excursions}
In this report we propose a method for identifying dynamic fluctuations in the functional connectivity between brain regions that, given their static connectivity, would be considered unexpected in a statistical sense. Using this approach we found that the resting-state is apparently punctuated by episodes during which many connections are unexpectedly strong or weak, episodes that we termed \textit{mass excursions}. Because these events tend to express a high proportion of the strongest functional connections (close to $\pm1$), they also produce more modular brain networks, with clear divisions between communities (i.e. sub-networks). This effect results in a dichotomy of brain states: In one state connections are unexpectedly strong and form tightly-bound and mutually-segregated communities. In the other state community structure is not entirely absent, though the overall distinctiveness and segregation of communities from one another has diminished. This conclusion is supported by converging lines of evidence, namely the statistically significant increase in $Q$ during mass-excursion episodes, as well as a concomitant reduction in the degeneracy of the modularity landscape.

We can interpret these results in several different ways. On one hand, the presence of well-defined, highly modular functional communities indicates that information may be selectively exchanged among or processed by their constituent brain regions. At the same time, communities are segregated from one another, which suggests that information may not be readily exchanged across community boundaries. Collectively, this leads us to hypothesize that the mass excursions, which correspond to highly modular functional network topologies, might represent periods of specialized information processing when information is less readily exchanged between modules. During time periods when few excursions take place, functional networks are less modular and feature communities that are not as well-defined; these periods might represent times during which information is more freely exchanged across community boundaries. This interpretation agrees with earlier results from \cite{liegeois2014cerebral} in which the authors reported oscillatory behavior between periods of high and low modularity, associated with a transient de-coupling of functional connections from the underlying anatomical network. It is also somewhat in line with earlier results \citep{zalesky2014time} reporting that functional networks flipped between periods of high and low global efficiency. We expect that further analysis will reveal that the high efficiency states reported by \cite{zalesky2014time} would correspond to the low modularity states observed here while the low efficiency states would be more closely related to high-modularity networks. This follows from the observation that greater global efficiency scores are associated with disordered (i.e. more random) networks, whereas higher modularity scores are associated with highly ordered (i.e. clustered) networks \citep{watts1998collective}. Future work can be directed to better understand the relationships of these states to one another.

Interestingly, our findings suggest that mass-excursion periods are typified by the dissociation of default mode regions dissociating from other systems and forming more distinct modules. One possible explanation for this observation is that instants during which default mode regions dissociate from other regions correspond to periods of introspective thought or mind-wandering, which may not require that system is integrated across multiple systems. This hypothesis is at least partly supported by previous studies that suggest that the default mode network's dynamic connectivity tracks daydreaming \citep{kucyi2014dynamic} and other drowsiness \citep{allen2012tracking}.

Finally, the intermittent occurrence of mass excursions is similar to other experimental results that suggest that the emergence of critical dynamics, canonical resting-state networks, and even BOLD activity is not a continuous process, but is driven by instantaneous events at the level of functional connections or in the activity profiles of brain regions, themselves \citep{tagliazucchi2012criticality, liu2013time, allan2015functional, karahanouglu2015transient}. Presently, however, the precise neurobiological relevance of the high- versus low-modularity states remains unclear. An intriguing lead in this regard is the apparent importance of somatosensory and visual networks, which participate disproportionately often in excursions and are, hence, more temporally variable than expected. Interestingly, functional connections between these systems have also been reported as being highly variable across scan sessions \citep{laumann2015functional}. Future work is needed to determine if mass excursions are driven by dynamic events involving these networks or another subset of connections.

\textit{red}{It should be noted that modularity scores, $Q$, are in general, biased by a network's total connection weight. The modularity function used here has a particular bias wherein networks with strong, positive connection weights will tend to have higher modularity scores. In this light, one might anticipate our observation that episodes of high modularity tend to coincide with periods when dynamic connections are unexpectedly strong. Nonetheless, having many positive connections does not, by itself, guarantee a large $Q$; for that to occur, these positive connections also must be distributed in such a way that they cluster into communities.}

\subsection*{Possible improvements}
\begin{figure}[ht]
\centering
\includegraphics[width=\linewidth]{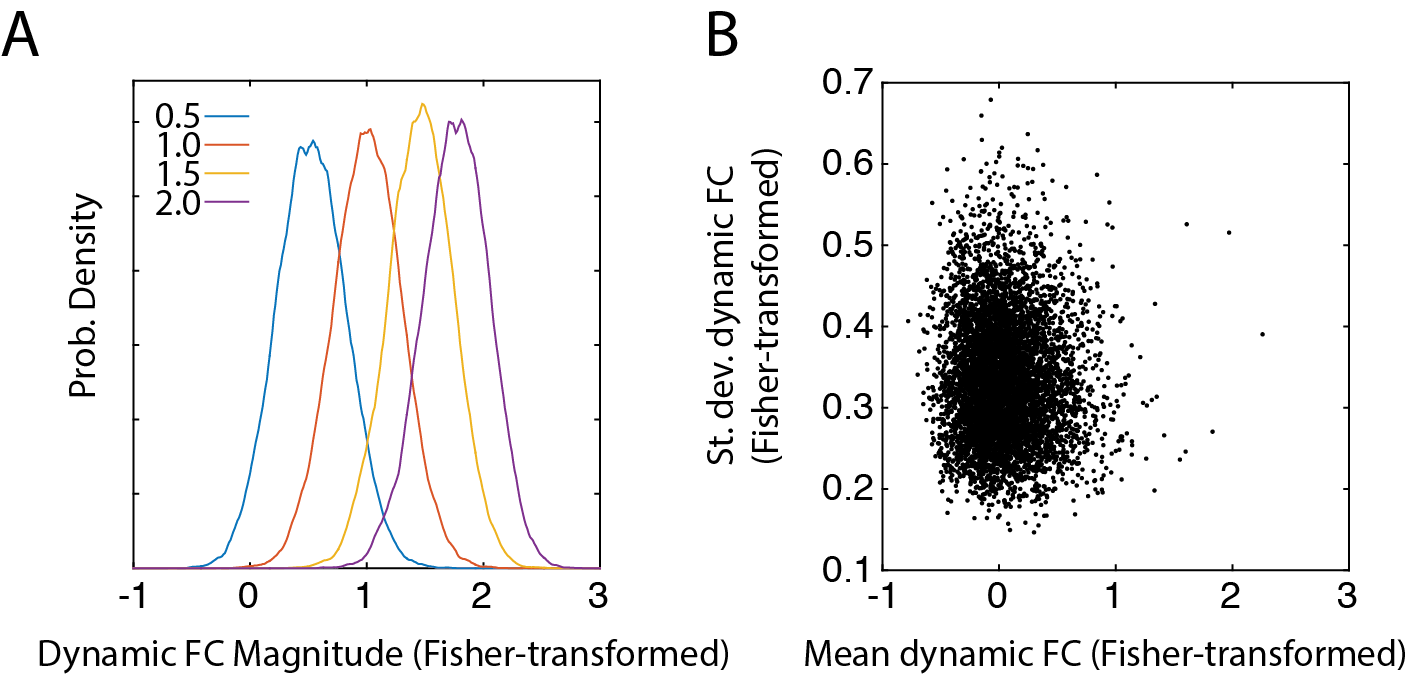}
\caption{We show the effects of Fisher transforming correlation coefficients on the range of dynamic fluctuations in Fisher-transformed coefficients. (A) Given four levels of static functional connectivity, we show the expected range of distribution Fisher-transformed dynamic functional connections. Note that despite the wide range of static connectivity magnitudes, the variances of distributions are similar. Contrast this with Figure 2 in the main text, where static connectivity magnitude played a massive role in shaping the range of dynamic fluctuations. (B) The relationship of the temporal mean and standard deviation for the Fisher-transformed functional connectivity matrix. Note that there is no systematic relationship. Contrast this panel with panel C in Figure 1 from the main text.}
\label{fig:sifig1}
\end{figure}

\begin{figure}[ht]
\centering
\includegraphics[width=\linewidth]{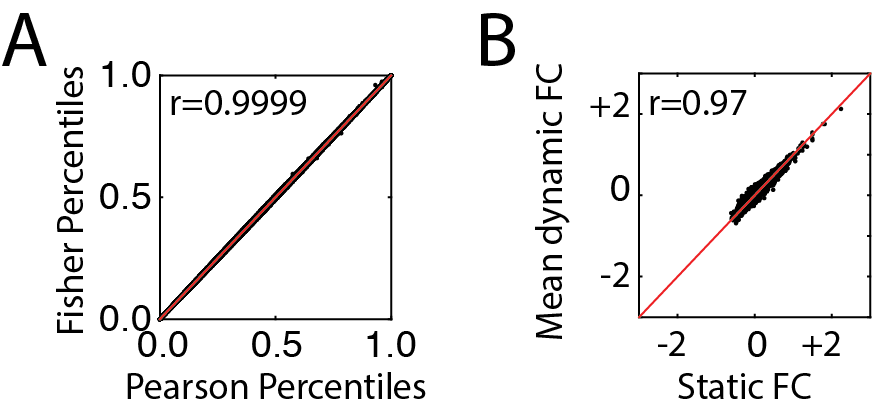}
\caption{Comparison of raw Pearson correlation coefficients with Fisher-transformed coefficients. (A) Because the Fisher transformation is monotonic (preserves rank) it does not change percentiles in any appreciable way. The correlation of the Fisher-transformed percentile matrix with the percentile matrix obtained from raw Pearson correlation coefficients is $r\approx1$. (B) We also calculated the correlation of Fisher-transformed static functional connectivity with the mean Fisher-transformed dynamic functional connectivity.}
\label{fig:sifig2}
\end{figure}

Our results are in line with earlier work by \citep{thompson2015mean} and \cite{zalesky2015towards}, indicating that a great deal of what we interpret as dynamic fluctuations in brain connectivity can be predicted on the basis of observed functional connectivity estimated over longer time scales. This predictability arises as a consequence of constraints imposed by the use of Pearson's correlation to estimate dynamic functional connectivity. For two variables (e.g. regional BOLD signals) to be strongly correlated over long intervals, it must be the case that, on average, they are also strongly correlated over shorter intervals. This relationship becomes stronger as the strength of correlation increases, which may explain why the least variable functional connections tend to be the strongest \citep{gonzalez2014spatial}. One possible strategy to help mitigate such issues is to Fisher transform the Pearson correlation coefficients \citep{fisher1915frequency}. The Fisher transformation effectively improves the normality of correlation coefficients so that they are no longer bound within the interval $[-1,1]$. Thus, as the strength of correlation between two variables increases, their variability stays approximately the same (See Figure \ref{fig:sifig1}). However, because the Fisher transform is monotonic the conditional dynamic functional connections estimated from Fisher-transformed data are basically unchanged from those estimated using Pearson's correlation coefficients (Figure \ref{fig:sifig2}).

\begin{figure}[ht]
\centering
\includegraphics[width=\linewidth]{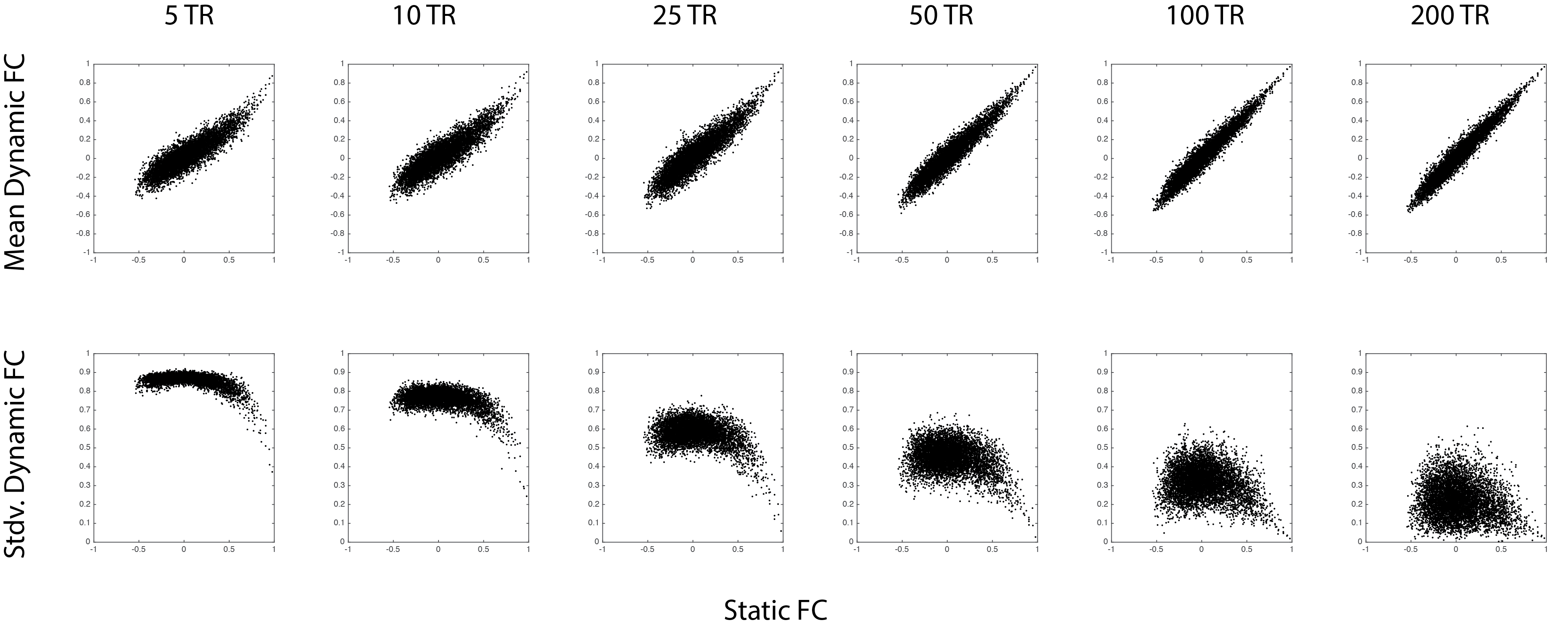}
\caption{Effect of non-overlapping windows on the relationship between static and dynamic FC. Each column displays the results using non-overlapping windows of different size (shown in terms of TRs/window). The top row shows the relationship between  static functional connectivity (x-axis) and mean dynamic functional connectivity (y-axis). In all cases, there is a clear nearly-linear relationship. In the second row, we plot the correspondence between static functional connectivity (x-axis) versus the standard deviation dynamic functional connectivity (y-axis). In all cases, the smallest standard deviations correspond to strongest static functional connections, indicating that these measures are not independent of one another.}
\label{fig:sifig3}
\end{figure}

\begin{figure}[ht]
\centering
\includegraphics[width=\linewidth]{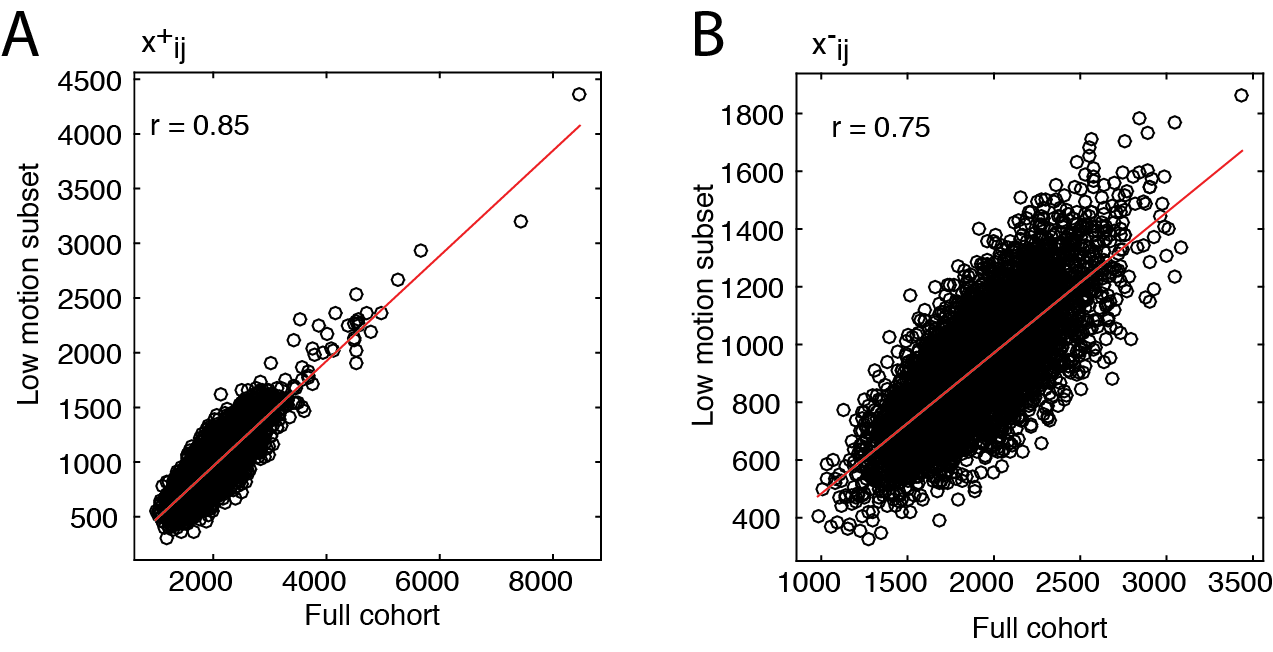}
\caption{Correlation of excursion count matrices generated from the full cohort against and from a sub-set of the participants whose excursion count time-series was not strongly correlated with framewise displacement. A) Scatterplot of $x^+_{ij}$ for both groups; B) Scatterplot of $x^-_{ij}$ for both groups.}
\label{fig:sifig4}
\end{figure}

There are other strategies for generating dynamic functional connectivity, including using non-overlapping windows. In this approach, the full fMRI BOLD time series are segmented into \textit{non-overlapping} intervals and a connectivity matrix created for each window as before. This approach has the obvious advantage that the window at time $t$ is independent of the window at time $t + 1$. In the supplementary section, we show, qualitatively, that the use of non-overlapping windows to estimate dynamic functional connectivity does not help reduce the dependence of dynamic functional connectivity magnitude on the static correlation magnitude. For a single subject we window the 885 TRs into non-overlapping windows of 5, 10, 25, 50, 100, and 200 TRs. We show that in all cases, there is a strong correlation between the mean dynamic functional with the static connectivity and that, to varying degrees, static connectivity still constrains the shape of the standard deviation in dynamic functional connectivity (Figure \ref{fig:sifig3}).

Our results were based on a univariate analysis of fluctuations in dynamic functional connections. Specifically, in estimating the number of connections that were unexpectedly strong or weak at any time point, we considered each connection independently. Of course, due to the correlative nature of the connection weights, functional connections are very often not independent of one another. Future work could include explicitly multivariate methods, as in \cite{leonardi2013principal}, where the authors identified connection-level components with spatial distributions reminiscent of known functional systems that collectively co-varied together across time. Importantly, this component-style analysis included a step in which time-averaged connectivity levels were subtracted from each connection's time course prior to estimating each component's temporal profile. This reveals patterns of temporal fluctuations of functional connections while discounting the baseline level of dynamic functional connectivity, which, as we and others show, is constrained by the level of static connectivity.

\subsection*{Limitations}
In this report we argue that the magnitude of static functional connectivity plays an important role in constraining expectations about dynamic fluctuations. We cite some practical examples using fMRI BOLD times series to motivate our claim. It is important to note, however, that in all but the most trivial examples (i.e. correlations of $\pm1$), it is mathematically possible for a dynamic fluctuation to take on any possible correlation value between -1 and 1. While this may be true, we assert that the probability of observing such fluctuations is not uniform, and that the shape of such a probability distribution is determined in part by the static connectivity magnitude (as well as other factors, such as window length, whether windows overlap or not, the power spectrum of the time series, etc.). It is also worth noting that our approach aims to estimate this distribution as closely as possible by generating many surrogate time series samples that have many properties in common with the empirical fMRI BOLD time series.

Our results are limited, in part, by the length of fMRI scan sessions. For example, we estimated the static functional connectivity for each participant based on 885 observations (time points or TRs, corresponding to approximately 9.37 minutes), and as a result of the finite sample size, consequently may not know the \textit{true} (i.e. very long term) values of static connectivity between pairs of regions \citep{laumann2015functional}.

In this paper we focus on functional connectivity as calculated by Pearson's correlation coefficient, as it is the most frequently-used measure in the extant literature. It should be noted that there are many alternative functional connectivity metrics \citep{smith2011network}, some of which can be applied to time-varying networks as well. Determining whether static estimates of functional connectivity made from these alternative measures can be used to predict dynamic connectivity is beyond the scope of this report, though we suspect that a similar rationale applies to these alternative measures.

\subsection*{Concluding remarks}
We show that the use of correlation as a measure of dynamic functional connectivity implies a number statistical sequelae, for example that connectivity over longer time-scales constrains the expected dynamic fluctuations expected at shorter time scales. We propose a method for identifying functional connections that are unexpectedly strong or weak, applying this technique to fMRI BOLD data. From these data, we show that dynamic functional connectivity undergoes transient periods of both high and low modularity, which are driven by the over-/under-expression of unexpectedly strong/weak functional connections. The approach has practical significance, in that it represents a model-free framework for identifying the time and location of dynamic fluctuations that are, in a statistical sense, unexpected, and thus perhaps of greater neurobiological importance.

\subsection*{Acknowledgements}
The authors thank Timothy O. Lauman for reading an early draft of this manuscript. RB was  supported by the National Science Foundation/Integrative Graduate Education and Research Traineeship Training Program in the Dynamics of Brain-Body-Environment Systems at Indiana University supported by. OS was supported by NIH 1 R01 AT009036-01. MF was by a Uehara Memorial Foundation Postdoctoral Fellowship. XNZ was supported by the National Basic Research Program (973: 2015CB35170) and the Major Joint Fund for International Cooperation and Exchange of the National Natural Science Foundation (81220108014).

\clearpage

\section*{References}
\bibliography{time-varying-fc-biblio}

\clearpage

\beginsupplement

\end{document}